\documentclass[conference]{IEEEtran}
\usepackage[utf8]{inputenc}
\usepackage{amsmath,amssymb,amsfonts,amsthm}
\usepackage{algorithmic}
\usepackage{graphicx}
\usepackage{textcomp}
\usepackage{xcolor}
\usepackage{booktabs}
\usepackage{tabularx}
\usepackage{multirow}
\usepackage{hyperref}
\usepackage{cleveref}
\usepackage{float}
\usepackage{enumitem}
\usepackage[ruled,vlined]{algorithm2e}
\usepackage{orcidlink}

\hypersetup{
    colorlinks=true,
    linkcolor=blue,
    citecolor=blue,
    urlcolor=blue
}

\begin{document}

\title{Governed AI-Assisted Engineering: Graduated Human Oversight for Agentic Code Generation in Regulated Domains}

\author{
    \IEEEauthorblockN{
        Dr.\ Richard Kang~\orcidlink{0000-0001-8674-4825}
    }
    \IEEEauthorblockA{
        DoiT International\\
        richard@doit.com
    }
}

\maketitle

\begin{abstract}
The adoption of agentic AI coding systems --- where autonomous agents generate, review, test, and deploy code with minimal human intervention --- creates a governance challenge in regulated industries. Existing frameworks address AI-assisted development maturity or the productivity-reliability tension but offer no mechanism for calibrating human oversight intensity to regulatory impact. We present the Governed AI-Assisted Engineering (GAIE) framework, a three-tier graduated human oversight model for agentic code generation in regulated domains. GAIE introduces the Oversight Classification Model (OCM), a deterministic decision function that classifies code generation tasks by regulatory impact, customer proximity, reversibility, and data sensitivity to route them through one of three oversight tiers: human-in-the-loop (strategic functions), human-over-the-loop (customer-impacting), or automated-with-monitoring (internal). Each tier defines required evidence artifacts for compliance auditability. We map GAIE against the Bank of Thailand's 2025 AI risk-management policy and demonstrate cross-jurisdiction applicability to MAS (Singapore), NIST AI RMF, ISO/IEC 42001, and the EU AI Act. Evaluation through regulatory coverage analysis, comparative framework analysis, and analytical productivity modeling suggests that graduated oversight preserves 84--97\% of agentic coding velocity (central estimate: 91\%) while maintaining compliance evidence coverage for regulated functions. GAIE contributes a framework that explicitly bridges AI-assisted development maturity with regulatory governance through proportionate human oversight.
\end{abstract}


\noindent\textit{The views expressed in this paper are those of the authors in their individual capacity as researchers and do not represent the official position, policy, or interpretation of their respective employers or any regulatory authority. This work constitutes academic thought leadership and does not constitute compliance advice or an institutional compliance assessment.}

\section{Introduction}

\subsection{The Agentic Coding Paradigm Shift}

AI-assisted software development has undergone a qualitative transformation. Where early tools such as GitHub Copilot operated at the granularity of a single line or function --- suggesting completions within human-directed workflows --- modern agentic systems perform multi-step reasoning, autonomous planning, tool use, and long-horizon code generation with minimal human intervention~\cite{bhati2026,li2025}. Systems such as Claude Code, Devin, SWE-agent, OpenHands, and MetaGPT can receive a high-level task description and autonomously decompose it into subtasks, generate code across multiple files, write tests, debug failures, and prepare deployment artifacts~\cite{li2025,wang2024}.

The speed is unprecedented. Agentic coding systems produce in minutes what human developers produce in hours~\cite{luo2025}. Controlled studies report productivity gains of 20--56\% on well-scoped tasks~\cite{farrag2026}, though the most rigorous randomized controlled trial documents a 19\% slowdown for experienced developers and telemetry across large developer populations shows substantially more pull requests but far longer review times with flat delivery metrics~\cite{farrag2026}. This tension between generation speed and reliability assurance forms the core challenge this paper addresses.

\subsection{The Governance Challenge in Regulated Domains}

Regulated industries --- banking, insurance, healthcare, energy --- face supervisory expectations that extend to any AI system influencing decisions with customer or public impact. Financial regulators including the Bank of Thailand (BOT), the Monetary Authority of Singapore (MAS), the UK's Prudential Regulation Authority (PRA), and the US Office of the Comptroller of the Currency (OCC) require human oversight for high-impact AI-informed decisions~\cite{bot2025,mas2022,occ2024}.

When AI agents write code that \textit{governs} regulated functions --- credit approval logic, customer onboarding workflows, anti-money laundering screening rules --- the oversight question transfers from model output governance to development pipeline governance. The institution remains responsible for code running in production regardless of whether a human or an agent wrote it~\cite{treude2026}.

The Bank of Thailand's 2025 AI risk-management policy is explicit: human participation or human oversight is required whenever AI is used for strategic functions, defined to include credit approval, account opening approval, and approval of deposits, withdrawals, or transfers~\cite{bot2025}. The policy further mandates lifecycle-wide controls across data quality, model evaluation, explainability, and AI-specific cyber defenses.

\subsection{Scope Boundary}

Before proceeding, we establish what GAIE governs and what it does not:

\textbf{GAIE governs the software engineering process in which AI agents generate code.} Specifically, it addresses: the classification of code generation tasks by their regulatory risk profile; the level of human oversight applied during the generation-to-deployment lifecycle; and the evidence artifacts produced at each stage for regulatory auditability.

\textbf{GAIE does not replace:}
\begin{itemize}[nosep]
    \item \textbf{Model risk management (MRM)} for production AI/ML models that make customer-facing decisions --- these remain subject to OCC SR 11-7 / MAS model validation requirements
    \item \textbf{Cybersecurity review} --- GAIE's security scans are generation-time controls, not a substitute for enterprise security architecture review
    \item \textbf{SDLC controls} --- GAIE layers governance \textit{on top of} existing software development lifecycle controls; it does not replace them
    \item \textbf{Regulatory approval of customer-facing AI systems} --- governance of customer-facing AI falls under separate regulation (e.g., BOT \S4.3 Principle 5), not GAIE
\end{itemize}

This distinction is important: GAIE governs the \textit{development pipeline} (how code is written and deployed), not the \textit{production behavior} of AI systems that code may implement.

\subsection{The Problem: Uniform Oversight Does Not Scale}

Two naive approaches to governing agentic code generation both fail:

\textbf{``Review everything''} creates velocity collapse. Atlassian's deployment of human-in-the-loop software development agents found that high cost of human feedback and integration challenges significantly constrained the value proposition~\cite{takerngsaksiri2024,pasuksmit2025}. When every AI-generated artifact requires human review at traditional code-review speed, the productivity benefit of agentic coding is entirely consumed by review overhead --- creating what Farrag terms the ``Productivity-Reliability Paradox''~\cite{farrag2026}.

\textbf{``Review nothing''} creates regulatory non-compliance and unauditable code. Without governance, AI-generated code affecting strategic functions reaches production without evidence that human oversight occurred --- a direct violation of regulatory expectations~\cite{bot2025,mas2022}.

The analogy to automotive autonomy is instructive. SAE J3016~\cite{sae2021} defines six levels of driving automation, each with different human attention requirements. The automotive industry does not debate ``should cars be autonomous or not?'' --- it asks ``what level of autonomy is appropriate for this driving context?'' We apply the same principle to code generation: not ``should agents code autonomously?'' but ``what level of oversight is appropriate for this code's regulatory context?''

\subsection{Contributions}

This paper makes the following contributions:

\begin{enumerate}[nosep]
    \item \textbf{The Governed AI-Assisted Engineering (GAIE) framework} --- a three-tier graduated human oversight model for agentic code generation that preserves velocity for low-risk tasks while enforcing human accountability for regulated functions.
    \item \textbf{The Oversight Classification Model (OCM)} --- a formal, deterministic decision function that routes code generation tasks to appropriate oversight tiers based on four risk dimensions. We establish key properties (monotonicity, fail-safety under correct or uncertain metadata, completeness) by construction.
    \item \textbf{A per-tier evidence artifact model} --- specifying what compliance evidence each tier produces, enabling regulatory auditability as a byproduct of development rather than a separate workstream.
    \item \textbf{Regulatory mapping and cross-jurisdiction analysis} --- providing a traceability mapping from GAIE components to selected regulatory control expectations across five frameworks.
    \item \textbf{Multi-method evaluation} --- including regulatory coverage analysis, comparative framework analysis, and analytical productivity impact estimation with sensitivity analysis.
\end{enumerate}

\subsection{Running Example}

Throughout this paper, we use a fictional running example drawn from a generic regulated bank implementing agentic coding. This example is illustrative and does not represent any specific institution.

\begin{quote}
\textbf{Scenario (illustrative --- not based on any specific institution):} A composite regulated commercial bank adopts an agentic coding system for its engineering teams. The bank's codebase includes: (a) credit decisioning microservices that determine loan approvals, (b) customer-facing mobile banking APIs, (c) internal developer tooling and CI/CD pipelines. The bank must comply with national AI risk management regulations that require human oversight for strategic functions.
\end{quote}

\subsection{Paper Organization}

Section~2 reviews background and related work. Section~3 presents the GAIE framework including the OCM, three-tier oversight model, and evidence artifact model. Section~4 maps GAIE to regulatory requirements across jurisdictions. Section~5 describes a reference implementation architecture. Section~6 presents evaluation results. Section~7 discusses implications, limitations, and future work. Section~8 concludes.

\section{Background and Related Work}

\subsection{Evolution of AI-Assisted Software Development}

\begin{figure*}[htbp]
    \centering
    \includegraphics[width=\textwidth]{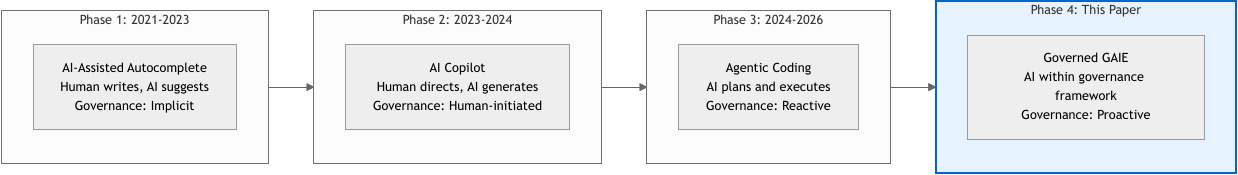}
    \includegraphics[width=0.5\textwidth]{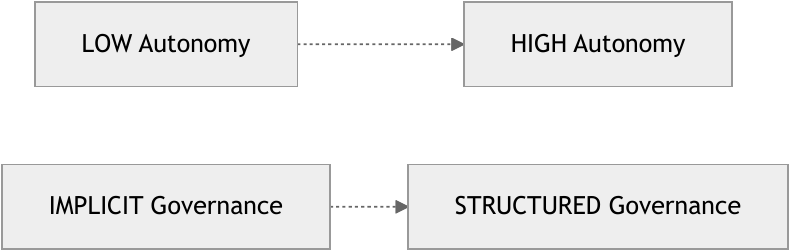}
    \caption{Evolution of AI-Assisted Software Development}
    \label{fig:evolution}
\end{figure*}

We identify four phases in the evolution of AI-assisted software development, depicted in \Cref{fig:evolution}.

\textbf{Phase 1 --- AI-assisted autocomplete} (2021--2023). Tools such as GitHub Copilot~\cite{copilot2021} and TabNine suggest single-line or function-level completions within human-directed workflows. Governance is implicit --- every suggestion requires explicit human acceptance. Productivity studies report 55\% faster task completion for basic coding tasks~\cite{peng2023}.

\textbf{Phase 2 --- AI copilot} (2023--2024). Conversational AI assistants integrated into IDEs generate multi-line code blocks in response to natural-language prompts. Governance remains human-initiated~\cite{chatgpt2022}.

\textbf{Phase 3 --- Agentic coding} (2024--2026). Autonomous systems decompose high-level tasks, make multi-step decisions, use tools, and produce complete implementations~\cite{bhati2026,li2025,wang2024}. SWE-bench evaluations show these systems resolving 20--70\% of real GitHub issues autonomously~\cite{jimenez2023}.

\textbf{Phase 4 --- Governed AI-assisted engineering} (this paper). Agentic systems operate within a governance framework that applies proportionate oversight calibrated to the regulatory impact of each task.

\subsection{Human-in-the-Loop in Software Engineering}

CentaurEval~\cite{luo2025} introduces a benchmark for measuring the value of human presence in agentic coding, finding that human reasoning combined with AI efficiency outperforms either alone on complex tasks. Takerngsaksiri et al.~\cite{takerngsaksiri2024} and Pasuksmit et al.~\cite{pasuksmit2025} report on Atlassian's deployment of HITL agents, identifying high cost and the need for structured interaction protocols. Mozannar et al.~\cite{mozannar2025} present Magentic-UI with co-planning and action-guarding patterns. Together these works establish that human presence adds value but is costly when applied uniformly --- motivating oversight that is calibrated rather than constant.

\subsection{Maturity Models for AI in Software Engineering}

Anderson~\cite{anderson2026} presents the AI Codebase Maturity Model (ACMM), a six-level framework (Level 0--5) describing how codebases evolve from basic AI-assisted coding to fully autonomous systems. ACMM describes \textit{what level of AI assistance exists} but not \textit{how to govern it proportionately}. Zohaib et al.~\cite{zohaib2025} present a layered maturity model for agentic AI in 6G software businesses. The CMMI lineage~\cite{cmmi2010} provides precedent for staged maturity models.

\subsection{The Productivity-Reliability Tension}

Farrag~\cite{farrag2026} identifies the ``Productivity-Reliability Paradox'': AI coding shows 20--56\% productivity gains on well-scoped tasks but 19\% slowdown for experienced developers, with 91\% longer review times. Farrag proposes specification-driven governance as the resolution --- applying uniformly regardless of regulatory risk. Our work observes that the paradox is an artifact of \textit{uniform} governance applied to \textit{heterogeneous risk}.

\subsection{Graduated Autonomy in Safety-Critical Domains}

The principle of graduated autonomy has precedent: SAE J3016~\cite{sae2021} (automotive, L0--L5), Parasuraman et al.~\cite{parasuraman2000} (aviation), Zabolotnii et al.~\cite{zabolotnii2026} (clinical AI staged autonomy), and NUREG~\cite{nureg2020} (nuclear safety classification). These domains share: \textbf{oversight intensity should be proportionate to consequence severity}. GAIE applies this principle to software engineering.

\subsection{Governance Frameworks for Agentic AI in Software Engineering}

Several recent works address fragments: Ait et al.~\cite{ait2025} (governance DSL), Treude et al.~\cite{treude2026} (accountability gaps), Hasanli~\cite{hasanli2026} (TDD governance), Boyuan et al.~\cite{boyuan2026} (dual-helix governance), Zietsman et al.~\cite{zietsman2026} (governance prompt quality), and Alenezi et al.~\cite{alenezi2026} (SE reorientation). None provides a comprehensive tiered oversight framework calibrated to regulatory requirements.

\subsection{Regulatory Frameworks for AI}

The Bank of Thailand~\cite{bot2025} requires human participation for strategic functions, lifecycle governance, and FEAT principles. MAS FEAT~\cite{mas2022}, NIST AI RMF~\cite{nist2023}, ISO/IEC 42001~\cite{iso42001}, BIS~\cite{bis2024}, the FSB~\cite{fsb2024}, and the EU AI Act~\cite{euaiact2024} all mandate \textit{proportionate} oversight but do not specify \textit{how} to implement proportionality in development workflows. GAIE fills this gap.

\subsection{Summary and Gap Statement}

No existing work combines: (a) graduated human oversight, (b) calibrated to regulatory impact classification, (c) for agentic code generation specifically, (d) with per-tier evidence artifacts, (e) validated against real regulatory frameworks. \Cref{tab:related} summarizes.

\begin{table*}[htbp]
\centering
\caption{Related Work Comparison}
\label{tab:related}
\footnotesize
\begin{tabularx}{\textwidth}{lXXXXXX}
\toprule
\textbf{Dimension} & \textbf{ACMM} & \textbf{Farrag} & \textbf{Atlassian} & \textbf{Zabolotnii} & \textbf{SAE} & \textbf{GAIE} \\
\midrule
Graduated oversight & --- & --- & Uniform & Clinical & Driving & \textbf{By reg. impact} \\
Regulatory compliance & --- & --- & --- & Medical & Auto safety & \textbf{Financial, cross-juris.} \\
Evidence artifacts & --- & Specs & --- & Evidence & --- & \textbf{Per-tier} \\
Classification model & Feedback & --- & --- & Risk & Context & \textbf{OCM: 4 dim.} \\
Formal properties & --- & --- & --- & --- & --- & \textbf{3 theorems} \\
\bottomrule
\end{tabularx}
\end{table*}

\section{The Governed AI-Assisted Engineering (GAIE) Framework}

\subsection{Framework Overview}

\begin{figure*}[htbp]
    \centering
    \includegraphics[height=0.88\textheight,keepaspectratio]{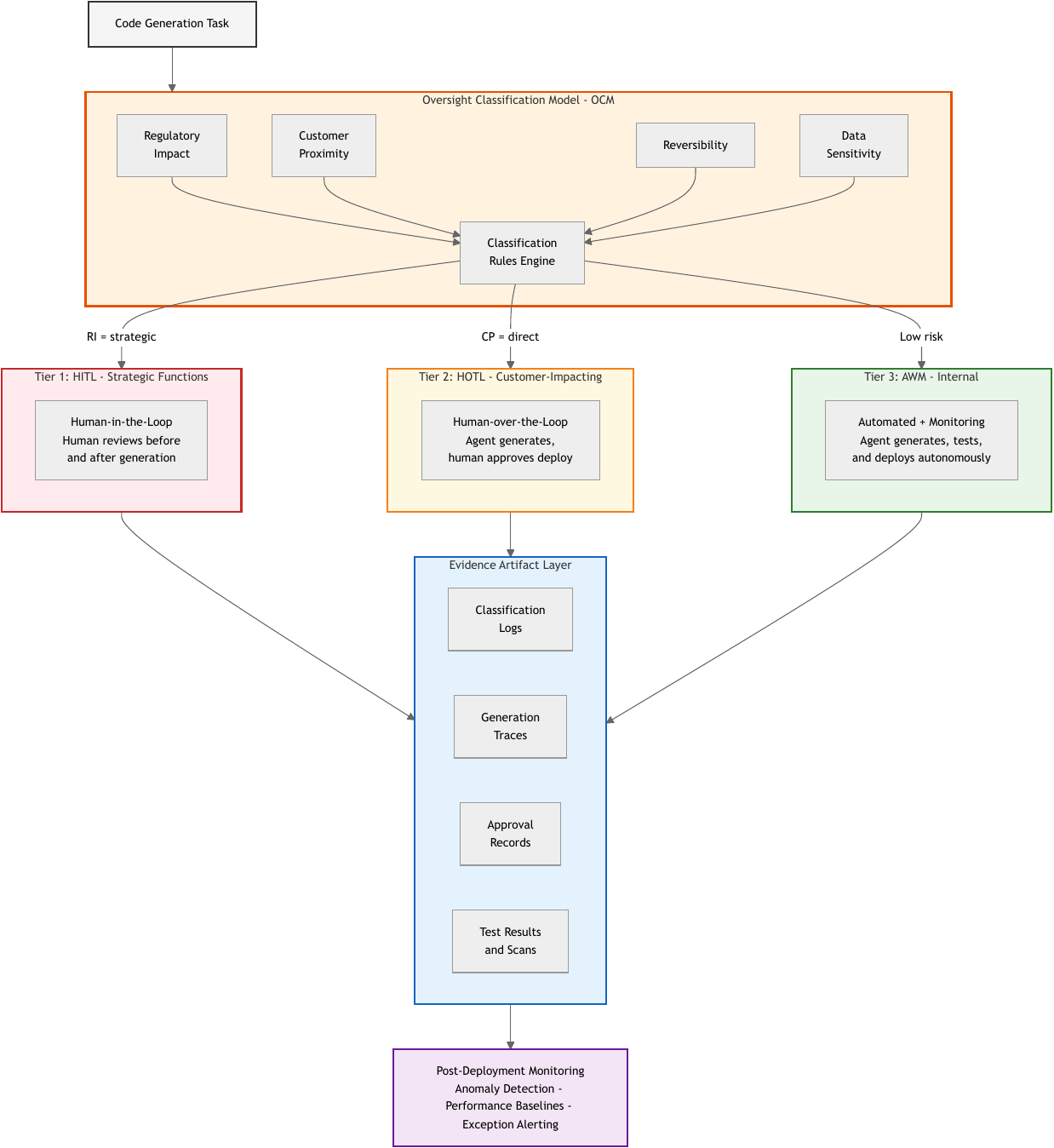}
    \caption{GAIE Framework Architecture Overview}
    \label{fig:architecture}
\end{figure*}

\Cref{fig:architecture} presents the GAIE framework architecture. The framework operates as a governance layer between task intake and code deployment, routing each code generation task through the appropriate oversight tier based on its risk classification.

\subsection{Design Principles}

GAIE is built on five design principles:

\begin{description}[nosep]
    \item[P1. Proportionality.] Oversight intensity scales with regulatory impact.
    \item[P2. Evidence-by-design.] Every oversight event produces auditable artifacts automatically.
    \item[P3. Fail-safe default.] When classification is uncertain, the OCM escalates to a higher oversight tier, never a lower one.
    \item[P4. Separation of generation and approval.] The agent that generates code cannot approve its own deployment.
    \item[P5. Regulatory defensibility.] Every framework element maps to at least one specific supervisory requirement.
\end{description}

\subsection{The Oversight Classification Model (OCM)}

\subsubsection{Formal Definition}

\textbf{Definition 1 (OCM).} The Oversight Classification Model is a total function:
\begin{equation}
    \text{OCM}: T \rightarrow \{\text{Tier1}, \text{Tier2}, \text{Tier3}\}
\end{equation}
where $T$ is the set of all code generation tasks.

\textbf{Definition 2 (Risk Feature Space).} Each task $t \in T$ is characterized by a risk feature vector:
\begin{equation}
    \varphi(t) = (\text{RI}(t), \text{CP}(t), \text{RV}(t), \text{DS}(t))
\end{equation}

\begin{table*}[htbp]
\centering
\caption{OCM Risk Dimensions}
\label{tab:dimensions}
\footnotesize
\begin{tabularx}{\textwidth}{llX}
\toprule
\textbf{Dimension} & \textbf{Domain} & \textbf{Definition} \\
\midrule
Regulatory Impact (RI) & \{strategic, non-strategic\} & Whether the code affects functions a regulator classifies as requiring human oversight \\
Customer Proximity (CP) & \{direct, indirect, internal\} & Whether the code directly affects customer-facing systems, indirectly supports them, or is purely internal \\
Reversibility (RV) & \{irreversible, partial, full\} & Whether deployed code can be rolled back without customer harm \\
Data Sensitivity (DS) & \{personal, business, public\} & Whether code generation context includes regulated personal data \\
\bottomrule
\end{tabularx}
\end{table*}

\subsubsection{Classification Decision Function}

\begin{figure*}[htbp]
    \centering
    \includegraphics[height=0.88\textheight,keepaspectratio]{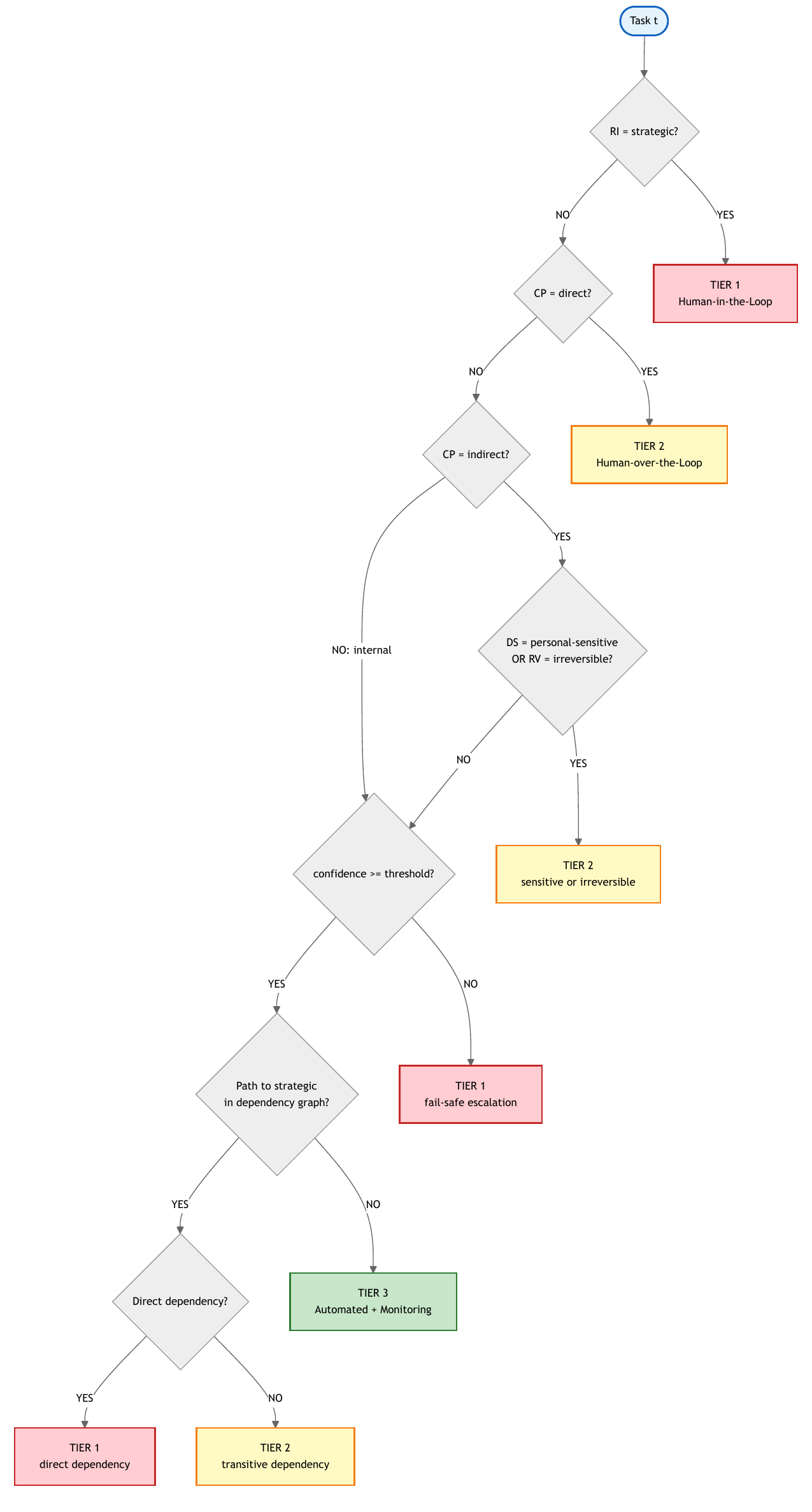}
    \caption{OCM Decision Tree}
    \label{fig:ocm}
\end{figure*}

\Cref{fig:ocm} presents the OCM classification decision tree.

\begin{algorithm}[t]
\SetAlgoLined
\KwIn{Task $t$, Registry $R$, Threshold $\theta$}
\KwOut{Tier assignment}
$ri \leftarrow \text{regulatory\_impact}(t, R)$\;
$cp \leftarrow \text{customer\_proximity}(t, \text{graph})$\;
$rv \leftarrow \text{reversibility}(t)$\;
$ds \leftarrow \text{data\_sensitivity}(t)$\;
$c \leftarrow \text{confidence}(ri, cp, rv, ds)$\;
\If{$ri = \text{STRATEGIC}$}{\Return Tier1}
\If{$cp = \text{DIRECT}$}{\Return Tier2}
\If{$cp = \text{INDIRECT} \wedge ds = \text{PERSONAL}$}{\Return Tier2}
\If{$cp = \text{INDIRECT} \wedge rv = \text{IRREVERSIBLE}$}{\Return Tier2}
\If{$c < \theta$}{\Return Tier1 \tcp*{fail-safe}}
\If{$\exists$ path in dependency\_graph$(t)$ to strategic}{
    \If{path\_length $= 1$}{\Return Tier1}
    \Else{\Return Tier2}
}
\Return Tier3\;
\caption{OCM Classification}
\label{alg:ocm}
\end{algorithm}

\subsubsection{Formal Properties}

\textbf{Theorem 1 (Monotonicity).} For any task $t$, increasing any risk dimension never decreases the assigned oversight tier.

\textit{Proof sketch.} The classification rules are ordered by decreasing tier assignment. If $\varphi(t_1) \leq \varphi(t_2)$ component-wise, then $\text{OCM}(t_1) \leq \text{OCM}(t_2)$. \qed

\textbf{Theorem 2 (Fail-Safety under correct or uncertain metadata).} If risk dimensions are correctly observed, or if uncertainty is detected below the confidence threshold, OCM does not under-classify tasks relative to their true risk profile.

\textit{Proof sketch.} Two cases: (a) Correct metadata: priority-ordered rules assign the tier corresponding to the highest-risk dimension present. (b) Uncertain metadata: $c < \theta \Rightarrow$ Tier1. \textit{Residual risk:} incorrect but confident metadata can cause misclassification. \qed

\textbf{Theorem 3 (Totality).} Every task $t \in T$ receives exactly one tier assignment.

\textit{Proof sketch.} The decision function terminates unconditionally with RETURN Tier3. The function is deterministic. \qed

\subsubsection{Running Example Application (Illustrative)}

\begin{table*}[htbp]
\centering
\caption{OCM Classification --- Illustrative Bank Scenario}
\label{tab:example}
\footnotesize
\begin{tabularx}{\textwidth}{Xllllc}
\toprule
\textbf{Task} & \textbf{RI} & \textbf{CP} & \textbf{RV} & \textbf{DS} & \textbf{Tier} \\
\midrule
Modify credit approval threshold & strategic & direct & irreversible & personal & \textbf{1} \\
Update mobile banking transfer UI & non-strat. & direct & full & business & \textbf{2} \\
Add field to internal logging & non-strat. & internal & full & public & \textbf{3} \\
Refactor AML screening rule & strategic & indirect & partial & personal & \textbf{1} \\
New CI/CD pipeline for test env & non-strat. & internal & full & public & \textbf{3} \\
API rate-limiting for customers & non-strat. & direct & full & business & \textbf{2} \\
\bottomrule
\end{tabularx}
\end{table*}

\subsection{Three-Tier Graduated Oversight}

\begin{figure*}[htbp]
    \centering
    \includegraphics[width=\textwidth,height=0.29\textheight,keepaspectratio]{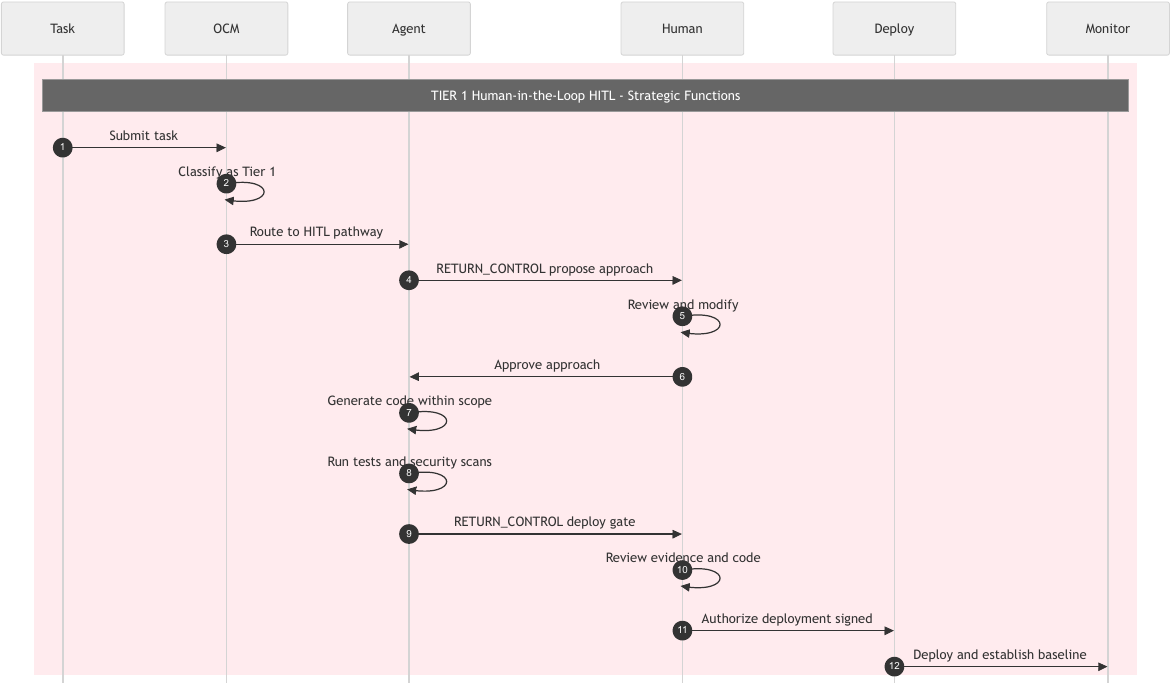}
    \vspace{0.3em}
    \includegraphics[width=\textwidth,height=0.29\textheight,keepaspectratio]{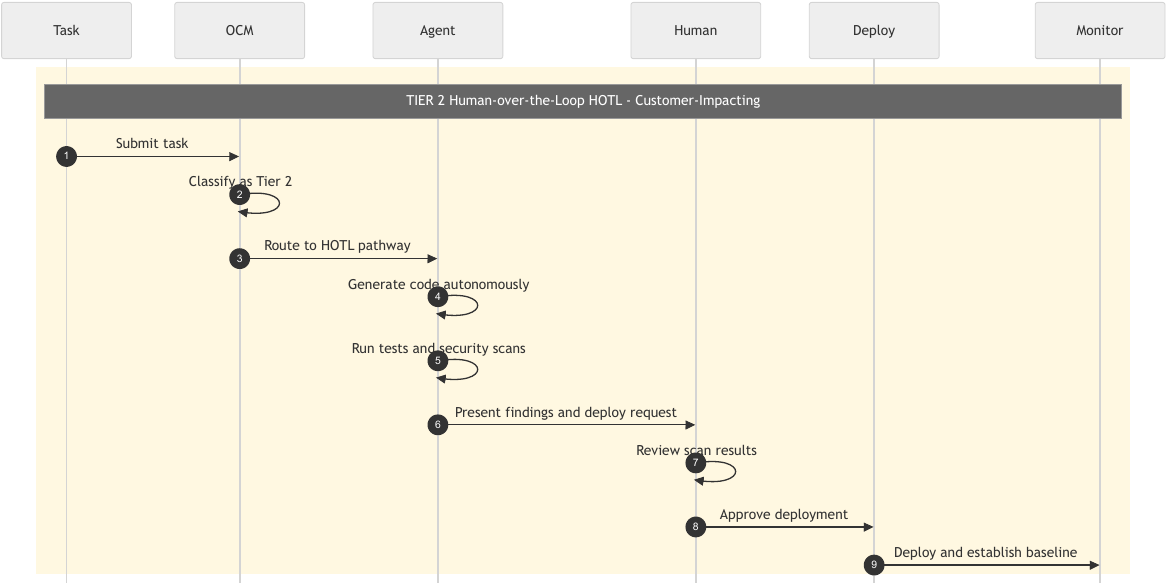}
    \vspace{0.3em}
    \includegraphics[width=\textwidth,height=0.29\textheight,keepaspectratio]{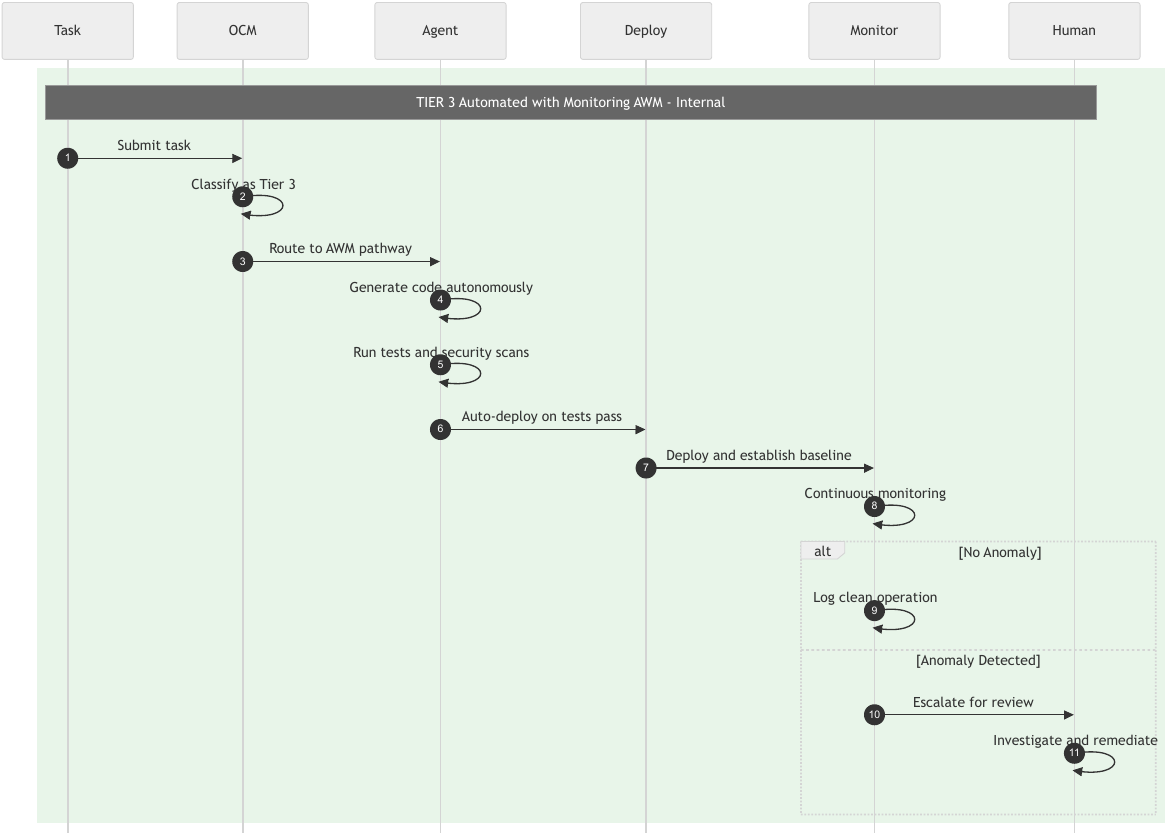}
    \caption{Three-Tier Oversight Model --- Sequence Diagrams}
    \label{fig:tiers}
\end{figure*}

\Cref{fig:tiers} illustrates the human involvement, agent behavior, and evidence flow at each tier.

\subsubsection{Tier 1: Human-in-the-Loop (HITL)}

\textbf{Trigger:} RI = strategic. The coding agent halts and escalates control (RETURN\_CONTROL). The human reviews the proposed approach, modifies if needed, and explicitly approves before generation proceeds. At deployment, the human again reviews all evidence and authorizes.

\textbf{Evidence produced:} Classification log, RETURN\_CONTROL event, human identity, review timestamps, approval decision with rationale, modification diff, deployment authorization with cryptographic signature.

\subsubsection{Tier 2: Human-over-the-Loop (HOTL)}

\textbf{Trigger:} CP = direct, OR (CP = indirect AND DS = personal), OR (CP = indirect AND RV = irreversible). The agent generates and tests autonomously; a human approves deployment.

\textbf{Evidence produced:} Classification log, generation trace, test results, security scans, human deployment approval.

\subsubsection{Tier 3: Automated with Monitoring (AWM)}

\textbf{Trigger:} All risk dimensions indicate low impact. Full autonomous pipeline with post-deployment monitoring and exception-based human escalation.

\textbf{Evidence produced:} Pipeline log, test results, deployment record, monitoring baseline, anomaly absence confirmation.

\subsection{Evidence Artifact Model}

\begin{table}[htbp]
\centering
\caption{Per-Tier Evidence Requirements}
\label{tab:evidence}
\footnotesize
\begin{tabular}{lcccc}
\toprule
\textbf{Artifact} & \textbf{T1} & \textbf{T2} & \textbf{T3} & \textbf{Rationale} \\
\midrule
OCM classification log & Req. & Req. & Req. & Auditability \\
Generation trace & Req. & Req. & Req. & Explainability \\
Human reviewer ID & Req. & Req. & --- & Accountability \\
Security scan results & Req. & Req. & Req. & Cyber controls \\
Test execution results & Req. & Req. & Req. & Evaluation \\
Deploy authorization & Signed & Signed & Auto & Approval \\
Monitoring record & Req. & Req. & Req. & Ongoing risk \\
Anomaly detection & Req. & Req. & Req. & Continuous mgmt \\
\bottomrule
\end{tabular}
\end{table}

\begin{figure*}[htbp]
    \centering
    \includegraphics[width=\textwidth]{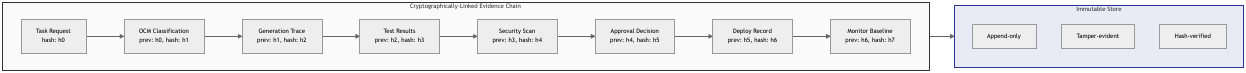}
    \caption{Evidence Chain Integrity Model}
    \label{fig:evidence}
\end{figure*}

\Cref{fig:evidence} shows the cryptographically-linked evidence chain.

\subsection{Framework Evolution: Tier Reclassification}

\begin{figure}[htbp]
    \centering
    \includegraphics[width=\columnwidth]{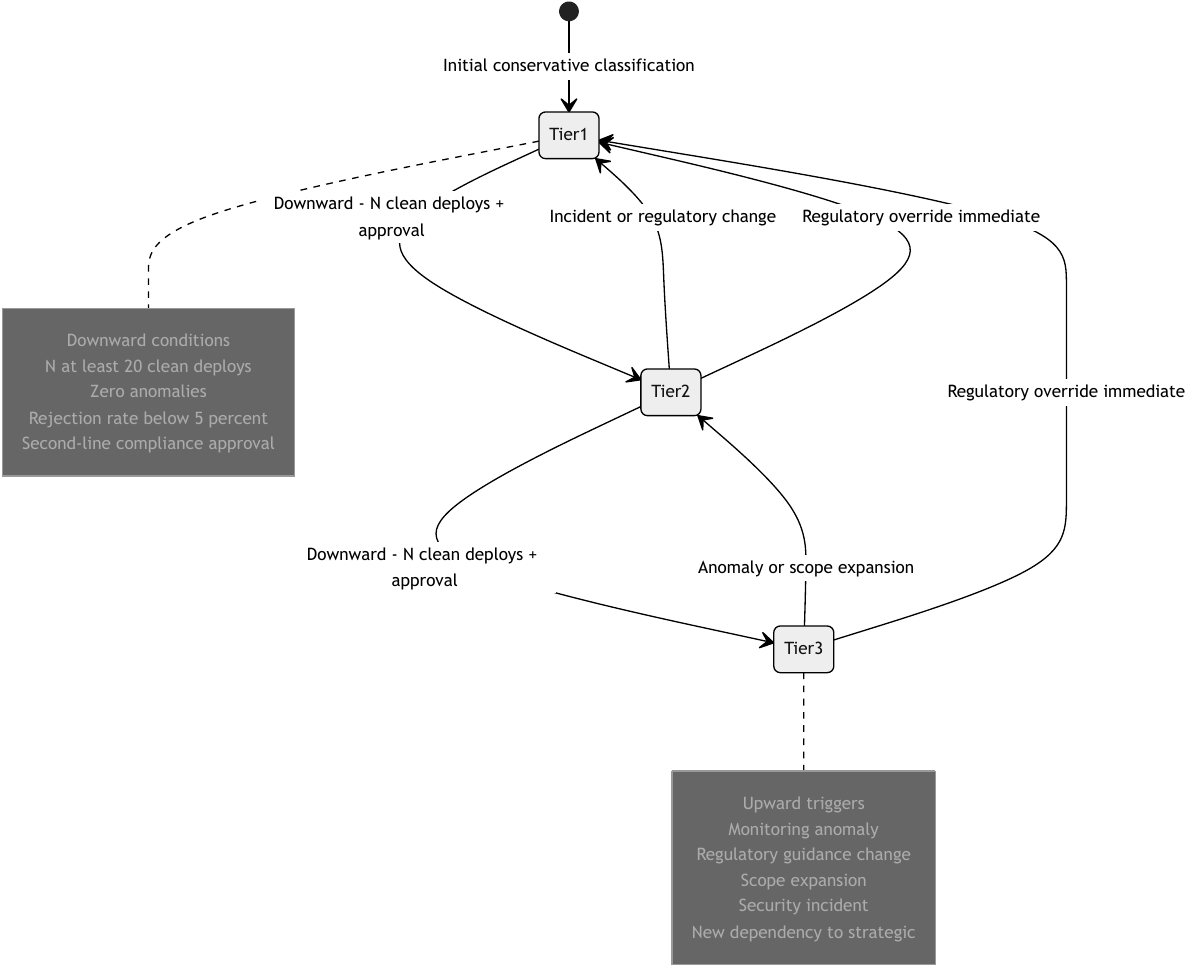}
    \caption{Tier Reclassification Lifecycle}
    \label{fig:reclass}
\end{figure}

\Cref{fig:reclass} depicts the tier reclassification lifecycle. Downward reclassification requires $N \geq 20$ consecutive clean deployments, zero anomalies, rejection rate below 5\%, and second-line compliance approval. Upward reclassification triggers immediately on monitoring anomaly, regulatory change, scope expansion, security incident, or new strategic dependency.

\subsection{Supporting Control Dimensions}

Three cross-cutting controls operate across all tiers:

\textbf{Data boundary enforcement.} Pre-transmission filtering (regex + NER-based PII detection), access scoping, and inference isolation.

\textbf{Prompt/response safety controls.} Aligned with OWASP LLM Top 10~\cite{owasp2025} and MITRE ATLAS~\cite{mitre2024}: prompt injection detection (LLM01), output validation for backdoors and exfiltration (ATLAS T0040--T0048), sandboxed execution, and supply chain validation (LLM05).

\textbf{Immutable audit trail.} Complete provenance chain from task request through deployment and monitoring.

\subsection{Threat Model for GAIE}

\Cref{fig:threats} presents the threat model. We identify four threat categories:

\begin{figure*}[htbp]
    \centering
    \includegraphics[width=\textwidth]{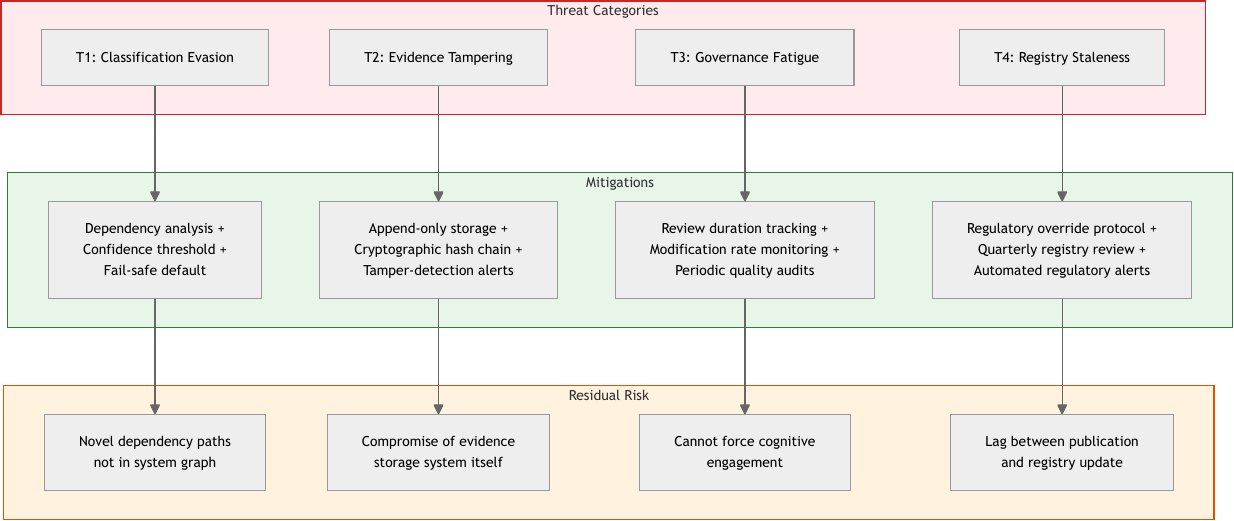}
    \caption{GAIE Threat Model}
    \label{fig:threats}
\end{figure*}

\begin{description}[nosep]
    \item[T1. Classification evasion.] Mitigation: dependency analysis, confidence threshold, fail-safe default.
    \item[T2. Evidence tampering.] Mitigation: append-only storage, cryptographic hash chain.
    \item[T3. Governance fatigue.] Mitigation: review duration tracking, modification rate monitoring.
    \item[T4. Registry staleness.] Mitigation: regulatory override protocol, quarterly review.
\end{description}

\section{Regulatory Mapping}

\subsection{Bank of Thailand AI Risk Management Policy (2025)}

We provide a traceability mapping from GAIE components to applicable BOT requirements (\Cref{tab:bot}).

\begin{table*}[htbp]
\centering
\caption{BOT Policy $\rightarrow$ GAIE Traceability Mapping}
\label{tab:bot}
\footnotesize
\begin{tabularx}{\textwidth}{XXX}
\toprule
\textbf{BOT Requirement} & \textbf{GAIE Component} & \textbf{Evidence} \\
\midrule
Human participation (\S4.3 P4) & OCM $\rightarrow$ Tier 1 + HITL & Classification + approval \\
Lifecycle governance (\S4.4) & Three-tier model & Per-phase artifacts \\
FEAT principles (\S4.3 P1--4) & OCM + evidence model & Audit trail \\
Data boundaries (\S4.4 Pt.2(1)) & Data boundary enforcement & Filtering logs \\
Testing (\S4.4 Pt.2(2)) & Gen-validation separation (P4) & Test results \\
AI cyber controls (\S4.4 Pt.2(3)) & Prompt/response safety & Attack detection logs \\
Explainability (\S4.3 P2) & Generation trace & Full trace \\
Three lines of defense & Agent/human/monitor & Role separation \\
\bottomrule
\end{tabularx}
\end{table*}

\textbf{Traceability assessment:} Based on the authors' reading of the publicly available BOT circular, GAIE's design addresses 9 of 10 applicable control domains. This represents academic analysis of published regulatory text and has not been validated by BOT examiners or any regulated institution's compliance function. It should not be construed as a compliance determination.

\subsection{Cross-Jurisdiction Applicability}

\begin{figure*}[htbp]
    \centering
    \includegraphics[width=0.9\textwidth]{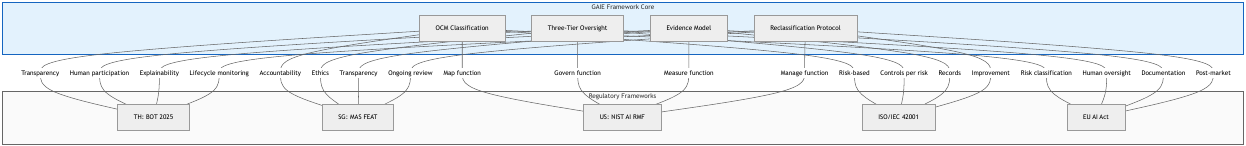}
    \caption{Cross-Jurisdiction Regulatory Mapping}
    \label{fig:crossjuris}
\end{figure*}

\Cref{fig:crossjuris} illustrates the mapping. \Cref{tab:crossjuris} details functional correspondence.

\begin{table*}[htbp]
\centering
\caption{Cross-Jurisdiction Traceability}
\label{tab:crossjuris}
\footnotesize
\begin{tabularx}{\textwidth}{lXXX}
\toprule
\textbf{Framework} & \textbf{Key Requirement} & \textbf{GAIE Mapping} & \textbf{Gap} \\
\midrule
MAS FEAT & FEAT principles & OCM transparency + human gates & Dev lifecycle \\
NIST AI RMF & Govern, Map, Measure, Manage & Four functions mapped & Full traceability \\
ISO/IEC 42001 & Risk-based AI mgmt system & OCM = risk-based approach & Clause 4 separate \\
EU AI Act & Human oversight (Art. 14) & Tier 1 for high-risk & Conformity not covered \\
OCC MRM & Model validation & Gen-validation separation & Complements, not replaces \\
\bottomrule
\end{tabularx}
\end{table*}

\textbf{Important caveat:} These mappings represent the authors' analysis and have not been validated by the respective regulatory authorities.

\subsection{Generalizability Beyond Financial Services}

The OCM's four risk dimensions are domain-agnostic. Only the regulatory function registry changes per domain. Examples in banking are drawn from publicly available regulatory guidance and are illustrative --- they do not represent any specific institution's classification decisions.

\section{Reference Implementation Architecture}

\subsection{Architecture Overview}

\begin{figure*}[htbp]
    \centering
    \includegraphics[height=0.82\textheight,keepaspectratio]{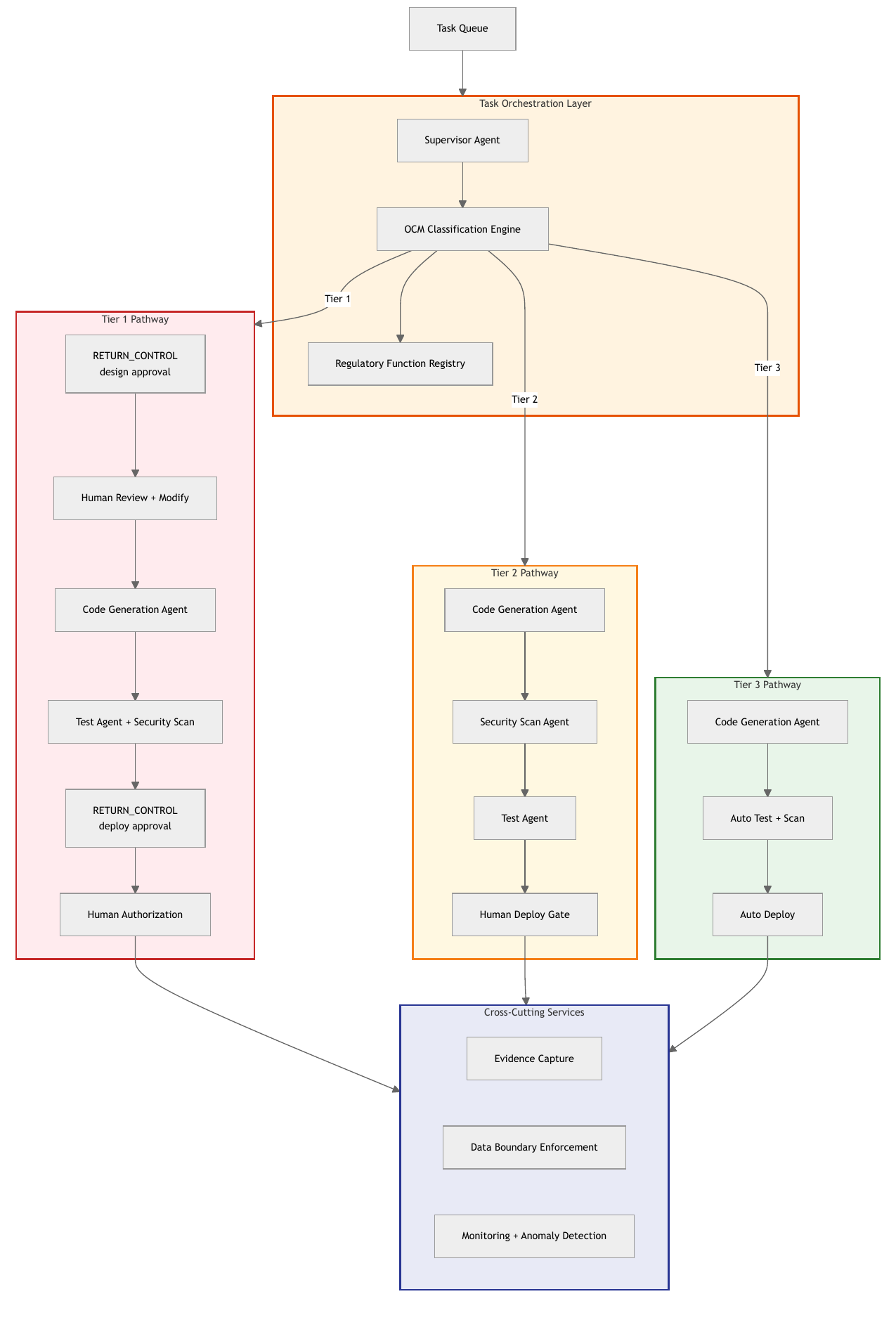}
    \caption{Reference Implementation Architecture}
    \label{fig:refarch}
\end{figure*}

\Cref{fig:refarch} presents the reference implementation architecture.

The architecture follows a multi-agent supervisor pattern with five subsystems: supervisor agent, collaborator agents, human interaction layer, evidence capture layer, and monitoring subsystem.

\subsection{Key Implementation Mechanisms}

\begin{table*}[htbp]
\centering
\caption{GAIE Concepts $\rightarrow$ Implementation Mechanisms}
\label{tab:mechanisms}
\footnotesize
\begin{tabularx}{\textwidth}{lXX}
\toprule
\textbf{Concept} & \textbf{Mechanism} & \textbf{Principle} \\
\midrule
OCM classification & Rule engine + registry & Deterministic routing \\
Tier 1 RETURN\_CONTROL & Agent escalation API & Halt + notify + await \\
Tier 2 deploy gate & CI/CD human approval & Block until authorized \\
Tier 3 auto deploy & Automated quality gates & Deploy on pass \\
Evidence capture & Invocation logging + events & Cryptographic chaining \\
Immutable storage & Append-only object store & Write-once, tamper-evident \\
Data boundaries & Pre-transmission filtering & No sensitive data to external APIs \\
Gen-validation separation & Distinct agent instances & Author $\neq$ tester \\
\bottomrule
\end{tabularx}
\end{table*}

\subsection{Recommended Adoption Sequence}

\begin{figure*}[htbp]
    \centering
    \includegraphics[width=\textwidth]{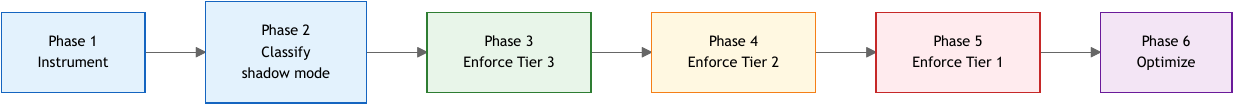}
    \caption{Phased Adoption Sequence}
    \label{fig:adoption}
\end{figure*}

\Cref{fig:adoption} presents the recommended phased adoption. Duration is institution-specific.

\begin{description}[nosep]
    \item[Phase 1: Instrument.] Deploy evidence capture and monitoring before granting agents deployment permissions.
    \item[Phase 2: Classify.] Run OCM in shadow mode. Validate against human judgment. Tune $\theta$.
    \item[Phase 3: Tier 3 enforcement.] Activate for internal tooling.
    \item[Phase 4: Tier 2 enforcement.] Extend to customer-impacting code with human deployment approval.
    \item[Phase 5: Tier 1 enforcement.] Extend HITL governance to strategic-function code.
    \item[Phase 6: Optimize.] Reclassify based on accumulated evidence.
\end{description}

\subsection{End-to-End GAIE Walkthrough}

\Cref{tab:walkthrough} traces three illustrative tasks through every GAIE stage. All task descriptions, system names, and parameters are fictional.

\begin{table*}[htbp]
\centering
\caption{End-to-End GAIE Walkthrough --- Three Illustrative Tasks}
\label{tab:walkthrough}
\footnotesize
\begin{tabularx}{\textwidth}{lXXX}
\toprule
\textbf{Stage} & \textbf{Tier 1: Credit} & \textbf{Tier 2: Mobile} & \textbf{Tier 3: CI/CD} \\
\midrule
Task & Lower loan threshold & Add date filter & Add retry logic \\
Risk & RI=strategic, CP=direct & CP=direct, RV=full & CP=internal, RV=full \\
Result & \textbf{Tier 1} & \textbf{Tier 2} & \textbf{Tier 3} \\
Generation & Human approves approach first & Autonomous & Autonomous \\
Human gate & Design + deploy & Deploy only & None \\
Monitoring & Enhanced 72h & Standard & Basic \\
Elapsed & $\sim$4 hours & $\sim$45 min & $\sim$8 min \\
\bottomrule
\end{tabularx}
\end{table*}

\section{Evaluation}

\subsection{Evaluation Methodology}

We evaluate GAIE through three complementary methods: (1) regulatory coverage analysis, (2) comparative framework analysis, and (3) analytical productivity impact estimation. This multi-method approach follows established practice for framework contributions~\cite{cmmi2010,kitchenham2002}. The evaluation is \textit{analytical} --- no production deployment data is available.

\subsection{Regulatory Coverage Analysis}

Of 18 enumerated BOT control requirements (6 organizational, 12 development/security), GAIE provides functional traceability to all 17 applicable requirements. The one non-applicable requirement (customer disclosure) is correctly excluded per the scope boundary (\S1.3).

\textbf{Important qualification:} Traceability does not equal compliance. This demonstrates that GAIE's design \textit{addresses} each requirement category; it does not constitute a compliance certification.

\subsection{Comparative Analysis}

\Cref{tab:comparative} shows feature comparison against related frameworks.

\begin{table*}[htbp]
\centering
\caption{Comparative Analysis Results}
\label{tab:comparative}
\footnotesize
\begin{tabularx}{\textwidth}{Xcccccc}
\toprule
\textbf{Dimension} & \textbf{ACMM} & \textbf{Farrag} & \textbf{Atl.} & \textbf{Zab.} & \textbf{GAIE} \\
\midrule
Regulatory compliance? & No & Partial & No & Med. & \textbf{Yes} \\
Productivity for low-risk? & N/A & Spec & No & N/A & \textbf{Yes} \\
Auditability evidence? & No & Spec & No & Yes & \textbf{Yes} \\
Proportionate oversight? & No & Uniform & Uniform & Grad. & \textbf{Yes} \\
Formal properties? & No & No & No & No & \textbf{Yes} \\
Fail-safe? & No & No & No & Partial & \textbf{Yes} \\
\bottomrule
\end{tabularx}
\end{table*}

\subsection{Analytical Productivity Impact Estimation}

Analytical modeling based on estimated task volume distributions (\Cref{tab:productivity}). These are analytical approximations, not empirical measurements.

\begin{table}[htbp]
\centering
\caption{Productivity Impact Model}
\label{tab:productivity}
\footnotesize
\begin{tabular}{lccc}
\toprule
\textbf{Parameter} & \textbf{Conserv.} & \textbf{Moderate} & \textbf{Optimistic} \\
\midrule
Tier 3 volume share & 60\% & 70\% & 80\% \\
Tier 2 volume share & 25\% & 20\% & 15\% \\
Tier 1 volume share & 15\% & 10\% & 5\% \\
Tier 3 velocity & 95\% & 98\% & 100\% \\
Tier 2 velocity & 80\% & 85\% & 90\% \\
Tier 1 velocity & 45\% & 55\% & 65\% \\
\midrule
\textbf{Weighted velocity} & \textbf{83.8\%} & \textbf{91.1\%} & \textbf{96.8\%} \\
\bottomrule
\end{tabular}
\end{table}

\textbf{Result: Analytical modeling suggests GAIE preserves 84--97\% of ungoverned agentic coding velocity} (central estimate: $\sim$91\%).

\begin{figure*}[htbp]
    \centering
    \includegraphics[width=0.48\textwidth]{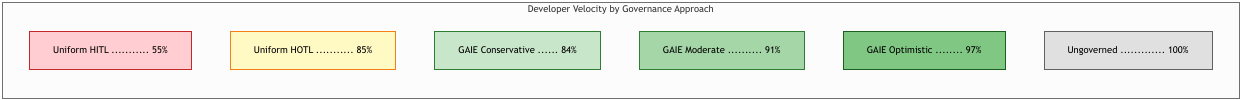}
    \hfill
    \includegraphics[width=0.48\textwidth]{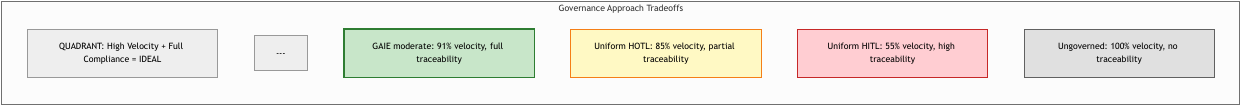}
    \caption{Productivity Impact --- GAIE vs. Uniform Oversight}
    \label{fig:productivity}
\end{figure*}

\subsection{Consistency with Empirical Baselines}

Our estimates are directionally consistent with: Farrag~\cite{farrag2026} ($\sim$81\% under uniform governance), Atlassian~\cite{takerngsaksiri2024,pasuksmit2025} (high HITL cost), and GitHub Copilot studies~\cite{peng2023} (55\% faster ungoverned). These remain \textit{analytical approximations}; empirical validation is future work.

\section{Discussion}

\subsection{Positioning Against Related Work}

GAIE complements ACMM (describes \textit{where}; GAIE prescribes \textit{how}), resolves Farrag's Paradox (by tiering rather than uniform governance), extends graduated autonomy to SE (from automotive, aviation, clinical domains), and operationalizes regulatory intent (translating principles into engineering processes).

\subsection{The Proportionality Principle}

Graduated oversight is superior on four grounds: \textbf{regulatory alignment} (aligns with the proportionality principle that regulators have established), \textbf{productivity evidence} (HITL adds value on complex judgment tasks, not routine generation~\cite{luo2025}), \textbf{governance quality} (concentrating attention prevents theatrical compliance), and \textbf{information theory} (graduated oversight allocates reviewer bandwidth to high-information decisions).

\subsection{Limitations and Threats to Validity}

\begin{description}[nosep]
    \item[L1.] Single jurisdiction for primary mapping; cross-jurisdiction lighter.
    \item[L2.] No production deployment data; estimates are analytical.
    \item[L3.] Classification boundary ambiguity at edges.
    \item[L4.] Expert validation pending (5--8 practitioners planned).
    \item[L5.] Technology coupling in reference architecture.
    \item[L6.] Adversarial robustness: confident but incorrect metadata is a failure mode.
    \item[L7.] Regulatory mappings not externally validated by authorities.
\end{description}

\subsection{Implications for Practice}

\textbf{For regulated institutions:} Adopt graduated oversight rather than the false dichotomy of ``review everything'' vs. ``ban AI coding.''

\textbf{For AI coding tool vendors:} Build RETURN\_CONTROL / escalation mechanisms as first-class features.

\textbf{For regulatory design (research observation):} Our analysis suggests that clarity of regulatory function definitions directly affects governance implementation efficiency. Where strategic function boundaries are clearly defined, institutions can calibrate oversight precisely.

\textbf{For researchers:} Empirical studies needed on OCM accuracy, tier calibration, false-escalation rates, and evidence artifact effectiveness.

\subsection{Future Work}

\begin{enumerate}[nosep]
    \item Empirical validation through production deployment
    \item Expert practitioner validation (Appendix B instrument)
    \item OCM automation via static analysis
    \item Machine-checked proofs (Coq, Lean)
    \item Extension to healthcare, insurance, energy
    \item Longitudinal reclassification study
    \item Integration with GRC platforms
    \item Multi-model governance
\end{enumerate}

\section{Conclusion}

Agentic AI code generation in regulated industries requires governance, but uniform oversight creates a productivity-reliability paradox. GAIE resolves this through graduated human oversight calibrated to regulatory impact.

The OCM routes tasks through three tiers based on regulatory impact, customer proximity, reversibility, and data sensitivity. Human-in-the-loop applies to strategic-function code (estimated 5--15\%). Human-over-the-loop applies to customer-impacting code (15--25\%). Automated-with-monitoring applies to internal tooling (60--80\%). Each tier produces defined evidence artifacts.

Evaluation suggests: (a) functional traceability to all applicable BOT lifecycle requirements, (b) differentiation from existing frameworks, (c) formal OCM properties (monotonicity, fail-safety, totality), and (d) 84--97\% velocity preservation (central: 91\%) vs. 45--65\% under uniform HITL.

GAIE contributes a framework that explicitly bridges AI-assisted development maturity with regulatory governance through proportionate human oversight --- making governance the enabler of velocity rather than its constraint.

\bibliographystyle{IEEEtran}
\bibliography{references}

\appendix

\section{BOT AI Policy Requirement Enumeration}

\textbf{Layer 1: Organizational Governance}
\begin{itemize}[nosep]
    \item A1. Board/senior management accountability
    \item A2. FEAT principles adoption
    \item A3. AI usage policy aligned with risk appetite
    \item A4. Three lines of defense
    \item A5. Human participation/oversight for strategic functions
    \item A6. Customer disclosure (not applicable to dev tools)
\end{itemize}

\textbf{Layer 2: Development and Security Controls}
\begin{itemize}[nosep]
    \item B1--B4. Data quality, assessment, leakage prevention, boundaries
    \item B5--B7. Model performance, testing, ongoing monitoring
    \item B8--B9. GenAI grounding, explainability
    \item B10--B12. Prompt filtering, attack testing, emerging threats
\end{itemize}

GAIE traceability: A1--A5, B1--B12. A6 not applicable (\S1.3).

\section{Expert Validation Instrument}

\textbf{Participant criteria:} Employed at regulated FI; role in engineering leadership, compliance, or security architecture; 3+ years experience; subject to tier-1 regulator.

\textbf{Quantitative (5-point Likert):}
\begin{enumerate}[nosep]
    \item GAIE addresses governance concerns for AI-assisted development
    \item Adoptable (as-is or adapted) for governing agentic coding
    \item Three-tier model appropriately calibrated
    \item Evidence artifacts sufficient for regulatory examinations
    \item OCM dimensions capture relevant risk factors
    \item Fail-safe default appropriate for risk tolerance
    \item Reclassification protocol provides adequate safeguards
\end{enumerate}

\textbf{Open-ended:} Gaps not addressed; tier boundary changes; comparison to current approach; adoption barriers; jurisdiction-specific gaps.

\end{document}